# A Peak Synchronization Measure for Multiple Signals

Rahul Biswas, Koulik Khamaru and Kaushik K. Majumdar, *Senior Member, IEEE*

*Abstract*—Peaks signify important events in a signal. In a pair of signals how peaks are occurring with mutual correspondence may offer us significant insights into the mutual interdependence between the two signals based on important events. In this work we proposed a novel synchronization measure between two signals, called *peak synchronization*, which measures the simultaneity of occurrence of peaks in the signals. We subsequently generalized it to more than two signals. We showed that our measure of synchronization is largely independent of the underlying parameter values. A time complexity analysis of the algorithm has also been presented. We applied the measure on intracranial EEG signals of epileptic patients and found that the enhanced synchronization during an epileptic seizure can be modeled better by the new peak synchronization measure than the classical amplitude correlation method.

## I. Introduction

SYNCHRONIZATION among different signals is an important feature of study for dynamical systems related to those signals [1] – [3]. Despite the predominant importance of synchronization there is no unique meaning assigned to it in physics [4] or in signal processing [5]. Intuitively, synchronization should give us a measure of interdependence or similarity between two signals. This has been studied by different methods for different applications such as amplitude correlation [6], mutual information [7], phase synchronization [8], [9] etc. One common trend across all these methods is to determine mutual dependence based on one particular feature uniformly across a segment of signals. Once a segment (or a window) is selected no effort is made to identify where in this segment that particular feature is more prominent or less prominent. Mutual dependence is calculated irrespective of the strength of the feature on different parts of the signal, whereas in reality, the strength of the features varies dynamically along individual signals.

In this work we propose a synchronization measure based on peaks in signals. In this paper we will be dealing only with upward or positive peaks. But all ideas developed can be extended to downward or negative peaks (troughs) in a straight forward manner. Here we will be concerned about peaks of signals alone, like in phase synchronization phase is considered alone, not amplitude etc. In many applications signal peaks are the focus of study [10] – [15]. In these applications a peak with much higher amplitude than the background signal contains information relatively more important than that in the background. If out of two signals $x(t)$ and $y(t)$, $x(t_n)$ is a peak and $y(t)$ is strongly dependent on $x(t)$ then $y(t_n)$ is expected to be a peak, where $t_n$ is a particular instance of $t$. If on the other hand $y(t)$ is weakly dependent on $x(t)$ then a peak may occur at $y(t_n + \tau)$ or at $y(t_n - \tau)$ for some $\tau > 0$, but not at $y(t_n)$ (we have taken weak dependence in this sense for this paper, it may have other interpretations elsewhere). For peak at $x(t_n)$ different weights are to be assigned for peaks at $y(t_n)$ and $y(t_n \pm \tau)$. Clearly, this scheme is capable of giving a measure of synchronization between $x(t)$ and $y(t)$. This type of a synchronization measure will be particularly useful for the signals in which peaks predominantly contain useful information. Biomedical signals form one such class.

So far no attempt has been made to define a synchronization measure among two or more signals based on occurrence of peaks or spikes in each of them. In this work we proposed one such measure for the first time. This is particularly important in neuroscience. Simultaneous occurrence of peaks across one or more signals collected from different parts of the brain is known as *event* [16], [17]. Many important events and artifacts appear in neural signals as peaks (spikes) and they appear simultaneously across multiple signals. This peak synchronization measure algorithm will be useful in detecting them. When multiple signals are coming from the same source, such as seismological signals from the same epicenter recorded at different geographical locations, they will have simultaneous or slightly delayed spikes, which can be measured for synchrony by this algorithm to ascertain that they are from the same source. In depth EEG recording of focal epileptic seizures spikes may appear in different focal channels with

Manuscript was submitted to the IEEE Transactions on Signal Processing on April 18, 2014, revised on June 15, 2014, and was accepted June 17, 2014. This work was supported in part by the Department of Science and Technology, Government of India, under grant SR/CSI/08/2009.

R. Biswas and K. Khamaru are with the Statistics and Mathematics Unit, Indian Statistical Institute, Kolkata 700108, India (e-mail: 1992.rahul@gmail.com; koulik123@gmail.com).

K. K. Majumdar, is with the Systems Science and Informatics Unit, Indian Statistical Institute, 8th Mile, Mysore Road, Bangalore 560059, India(e-mail: mkkaushik@hotmail.com).

some time lag. Yet the signals collected from different channels are known to be highly synchronous. We have shown in this work that this algorithm is an appropriate tool to model this kind of synchronization.

In the next section we present a detailed description of the proposed peak synchronization measure. In section 3 we show an application of this measure on epileptic iEEG signals and compare the peak synchronization with amplitude correlation. The paper concludes with a Conclusion section, which also incorporates future directions.

## II. PEAK SYNCHRONIZATION

### A. Peak Detection

Peaks are regions of much higher amplitude within a lower amplitude background in a signal. So far many different algorithms for peak detection have been developed, all of which have their respective pros and cons. Some commonly used simple algorithms for peak detection are the standard amplitude thresholding and finding every local maxima. There are also the traditional window-threshold techniques [10], [18–20]. Some of the other techniques include peak detection by wavelet transform based pattern matching [11], [12], [21–27], Hilbert transform [28], combining Hilbert and wavelet transform [29], artificial neural networks [30], [31], techniques using templates [32], [33], morphology filtering [34]–[36], Kalman filtering [37], Gabor filtering [38], Gaussian second derivative filtering [39], linear prediction analysis [40] and smoothed nonlinear energy operator [41].

Here we have used a threshold based peak detection algorithm [42], where the threshold is median + 2 times standard deviation within a window of the signal. The window is subsequently slided along the signal. The proposed peak synchronization measure does not depend on how peaks have been detected.

### B. Peak Comparison

Let $p_k$ be a sequence of signals, where $p_k(t)$ is the converted $t$th time point of the $k$th signal $S_k$. The conversion of time points of the original signals happens in the following manner:

$p_k(t) = 1$, if $S_k(t)$ is a peak,
$= 0$, otherwise.

We first consider two signals and then generalize the measure to more than two signals. Let us define the following quantities, whose significance will become clear subsequently.

$$f_k(t) = \sum_{j=-n}^{n} a_j p_k(t+j)$$
$$I_k(t) = 1, if \ p_k(t) = 0$$
$$= \frac{1}{2}, if \ p_k(t) = 1 \quad (1)$$

$k \in \{1,2\}$, $a_j$ are the weights, obtained by appropriately segmenting a probability density function into strips as shown in Fig. 1. For finding peak synchrony, one needs to take a symmetric probability density function, non-decreasing before 0 and non-increasing afterwards. Here let $2n+1$ be the number of strips having area not negligibly small (Fig. 1). This number is odd, because central strip + places for n number of peaks on either side of it.

The measure of peak synchronization between $S_1$ and $S_2$ at the $t^{th}$ time point, $\phi_{1,2}(t)$ is given by

$$\phi_{1,2}(t) = I_1(t)f_1(t)p_2(t) + I_2(t)f_2(t)p_1(t)$$
$$= \left(\sum_{k=1}^{2}(f_k I_k)(t)\right)\left(\sum_{k=1}^{2}p_k(t)\right) - \left(\sum_{k=1}^{2}(f_k I_k p_k)(t)\right)$$
(2)

For most practical purposes, one can use the normal density function with mean 0 and scale (standard deviation) 1 as the weight determination function. We will present a rigorous justification in support of this assertion later in this paper.

The intuitive idea behind the peak synchronization measure is, given any peak in $S_1$ at time point $t$ we look for a peak in $S_2$. If $S_2(t)$ is a peak we assign the highest weight ($a_0$) to the peak $S_2(t)$. The peaks on either sides of the peak $S_2(t)$ are assigned weights $a_j$'s (other than $a_0$). Weights are decreasing in magnitude with increasing distance of the peak from $S_2(t)$.

The weights are obtained by taking areas of strips, of equal width, under the probability density function. The width of the strips is determined by taking the area of the central strip equal to $a_0$ (Fig. 1). $f_k(t)$ captures the extent of presence of peaks around time point $t$. The role of $I_k(t)$ is to do an average when both $S_1(t)$ and $S_2(t)$ are peaks and $\phi_{1,2}(t)$ is the sum of measure of synchronization of peaks around $S_2(t)$ with the peak $S_1(t)$ and that of peaks around $S_1(t)$ with the peak $S_2(t)$.

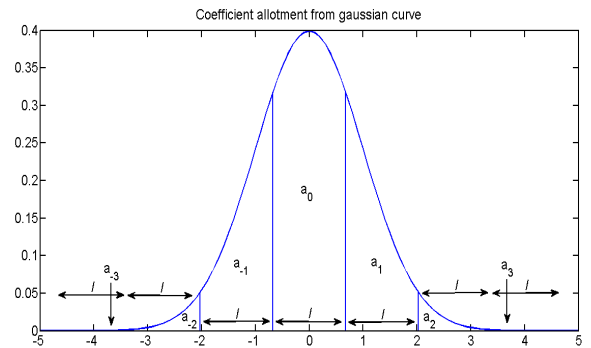

Fig. 1. In the above figure, the length $l$ is determined from the value assigned to $a_0$ (0.5 in the above figure). The other weights are determined by the area of adjacent strips of length $l$. The tail area rapidly decreases, the further we move from origin. Here, the tail area becomes insignificant (<10^-4) beyond $\pm 4.7$, and hence strips after $\pm 4.7$ are ignored. This determines the coefficient vector, which in this case, is $(a_{-3}, ..., a_0, ..., a_3)$.

## C. Generalization of the measure to r many signals

Generalizing the measure for r many signals ($r \geq 2$), we obtain the following formula, using the above definitions of $f_k$, $p_k$ and $I_k$ $\forall k \in \{1, 2, ..., r\}$ (equation (1))

$$\phi_{1,2,...,r}(t) = \frac{1}{\binom{r}{2}} \left[ \left( \sum_{k=1}^{r} (f_k I_k)(t) \right) \left( \sum_{k=1}^{r} p_k(t) \right) - \left( \sum_{k=1}^{r} (f_k I_k p_k)(t) \right) \right] = \frac{1}{\binom{r}{2}} \sum_{i=1}^{r-1} \sum_{j=i+1}^{r} \phi_{i,j}(t)$$
(3)

The factor $\frac{1}{\binom{r}{2}}$ ensures comparability of values obtained for clusters of signals of different sizes, by taking the average across all the pairs.

The proposition below shows that the formula for measure of peak synchrony for r many signals turns out to be the average of the of peak synchrony measure obtained from each of the $\binom{r}{2}$ pairs of signals (equation (3)), i.e.

$$\phi_{1,2,...,r}(t) = \frac{1}{\binom{r}{2}} \sum_{i=1}^{r-1} \sum_{j=i+1}^{r} \phi_{i,j}(t)$$

**Proposition 1.**
$$\left( \sum_{k=1}^{r} (f_k I_k)(t) \right) \left( \sum_{k=1}^{r} p_k(t) \right) - \left( \sum_{k=1}^{r} (f_k I_k p_k)(t) \right) = \sum_{i=1}^{r-1} \sum_{j=i+1}^{r} \phi_{i,j}(t)$$

*Proof.*

$$\left( \sum_{k=1}^{r} (f_k I_k)(t) \right) \left( \sum_{k=1}^{r} p_k(t) \right) - \left( \sum_{k=1}^{r} (f_k I_k p_k)(t) \right)$$
$$= \sum_{k=1}^{r} \sum_{j=1, j \neq k}^{r} (f_k I_k p_j)(t)$$
$$= \sum_{k=1}^{r-1} \sum_{j=k+1}^{r} (f_k I_k p_j)(t) + \sum_{k=2}^{r} \sum_{j=1}^{k-1} (f_k I_k p_j)(t)$$

The 2nd term above is equal to the same with the summations interchanged, i.e.

$$\sum_{k=2}^{r} \sum_{j=1}^{k-1} (f_k I_k p_j)(t)$$
$$= \sum_{j=1}^{r-1} \sum_{k=j+1}^{r} (f_k I_k p_j)(t)$$
$$= \sum_{k=1}^{r-1} \sum_{j=k+1}^{r} (f_j I_j p_k)(t)$$

So, the sum reduces to

$$\sum_{i=1}^{r-1} \sum_{j=i+1}^{r} ((f_i I_i p_j)(t) + (f_j I_j p_i)(t)) = \sum_{i=1}^{r-1} \sum_{j=i+1}^{r} \phi_{i,j}(t)$$

which proves the above proposition.

∎

Remark 1: Averaging $\phi_{1,2,...,r}(t)$ over time points in a certain interval $\{t_0, t_0 + 1, ..., t_0 + N - 1\}$ as following,

$$\Phi(1,2,...,r) = \frac{1}{N} \sum_{t=t_0}^{t_0+N-1} \phi_{1,2,...,r}(t)$$

leads to a single compound measure, indicating overall peak synchronization of the r signals in the specified interval. Calculating the same for the entire time duration, the compound measure can be used to sort groups of signals according to their overall peak synchronization, and particularly find the most peak synchronous group of signals, from a larger set of signals, or find the largest group whose measure of peak synchrony is greater than a threshold.

Remark 2: The choice of value of the central coefficient would depend on the signal being studied and needs to be fixed by the user. Intuitively, it tells us how likely the signals are peak-synchronized around a time point if both signals have a peak at that time point. One need not be too precise as the measure remains approximately invariant over a 0.2-neighborhood of central coefficients (See Fig 2). The peak synchrony measure becomes more spiky with increase in $a_0$. Also, the larger the value of $a_0$, the more is the sensitivity of the measure towards small changes. The value of peak synchrony measure becomes more stable if we choose smaller value of $a_0$. In fact, in absence of a specific choice of $a_0$, a value 0.5 for it would be safe for most purposes, in the sense that the curve of varying measure with time corresponding to $a_0 = 0.5$ is midway between the smoothest and the spikiest curves and the average measure lies midway between the highest possible and the lowest possible values. (See Fig 2, 3)

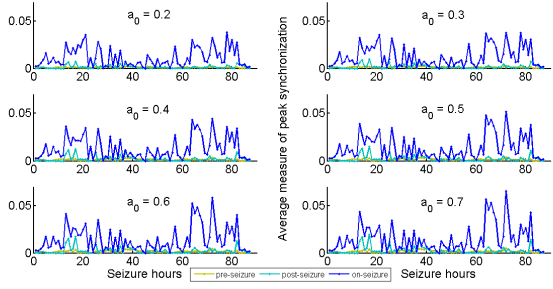

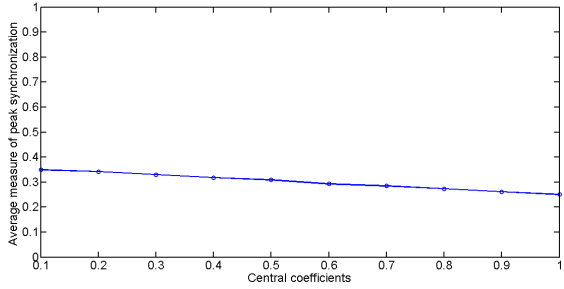

Fig. 2. The above plots depict 3 lines, points in which indicate the average measure of synchronization of neural signals from the 3 focal channels for pre (red), post (green) and during (blue) seizure for 87 recordings containing seizure, over a range of central coefficients. This figure shows the approximate invariance of the average measure of synchronization remains over seizure hours, for a wide range of central coefficients.

Fig. 3. The above figure records the variation of the average measure of peak synchrony of 3 uniform random binary signals over a range of central coefficients.

Remark 3: The measure is indeed non-negative, permutation invariant and is binless, i.e. the measure is independent of window length.

### D. Peak synchrony measure algorithm

*1) Algorithm for determination of weights*

Step1. Decide the choice of $a_0(>0)$ and the value of $\tau, 0 < \tau \ll 1$ to work with.

Step2. Find $x > 0$[1] such that $\int_{-x}^{x} f(t)dt = a_0$, where $f$ is any probability density function in the class described in equation (1). (For most practical cases, $f(t) = \frac{1}{\sqrt{2\pi}} e^{-\frac{t^2}{2}}$, the Gaussian function with mean 0 and scale parameter 1 can be used.)

Step3. Find smallest $n$ which satisfies

$$\int_{-(2n+1)x}^{(2n+1)x} f(t)dt \geq 1 - \tau$$

---
[1] For the class of functions considered x is unique. Although, if we had taken a larger class of functions with just f non-negative and $\int_{-\infty}^{\infty} f(x)dx = 1$, we could have worked with the minimum x such that $\int_{-x}^{x} f(t)dt = a_0$.

Step4. The value of $a_j$ is obtained by $a_j = \int_{(2j-1)x}^{(2j+1)x} f(t)dt$, $\forall j = \{-n, ..., n\}\setminus\{0\}$.

*2) Algorithm for peak synchronization measure calculation using the above found weights.*

Step1. Obtain the values of $a_j \forall j = \{-n, ..., n\}$, using the algorithm(1).

Step2. For $k = 1,2, ..., r$ and $t = 1,2, ..., N$.

- Calculate $f_k(t) = \sum_{j=-n}^{n} a_j p_k(t + j)$
- Calculate $I_k(t)$ from equation (1).

Step3. Calculate $\phi_{1,2,...,r}(t)$ given by

$$\phi_{1,2,...,r}(t) = \frac{1}{\binom{r}{2}} \left[ \left(\sum_{k=1}^{r} (f_k I_k)(t)\right)\left(\sum_{k=1}^{r} p_k(t)\right) - \left(\sum_{k=1}^{r} (f_k I_k p_k)(t)\right) \right]$$

### E. Computational Complexity of the Algorithm

Time complexity of the algorithm for measure calculation in D.2 is $O(r(r-1)(N-2n))$. In most practical cases, r even has a finite upper bound which does not depend on $N$, in which case, the time complexity is $O(N)$. Proof of both the above facts have been outlined in Appendix A.

### F. Invariance of weights with change in scale

Suppose one decides to use Gaussian function as a weight determination function. Now one natural question would be what should be a proper choice of the scale parameter $\sigma^2$. More generally, if one decides to use $f(x)$ as a weight determination function, then how the weight will change if $f_\sigma(x) = \frac{1}{\sigma} f\left(\frac{x}{\sigma}\right)$ is used instead of $f$. The proposition below shows that both $f$ and $f_\sigma$ give the same weight vectors. In particular using a Gaussian function with any scale parameter but same location parameter gives the same set of weight vectors.

**Proposition 2.**
The weights remain invariant with respect to change in scale of the weight determination function. In other words, suppose that in algorithm D, the value of $a_j$'s $\forall j \in \{-n, ..., n\}\setminus\{0\}$ were calculated using a function, $f(t), a_0 = a^*(>0)$ and $\tau = \tau^*, (0 < \tau^* \ll 1)$.
If one calculates the weights similarly, using same values of $a_0 = a^*$ and $\tau = \tau^*$, but using $f_\sigma(t) = \frac{1}{\sigma} f\left(\frac{t}{\sigma}\right)$, then denoting $a_j(\sigma), j = -n_\sigma, ..., n_\sigma$, to be the new weights,
1. $n_\sigma = n$, i.e. number of weights remain same.
2. $a_j(\sigma) = a_j \forall j \in \{-n, ..., n\}$, i.e. the value of the weights remain same.

*Proof.*
Firstly, $a_0(\sigma) = a_0 = a^*$.

Using the above, we will show that, if $x_\sigma, x \ (> 0)$ are such that $\frac{1}{\sigma} \int_{-x_\sigma}^{x_\sigma} f\left(\frac{t}{\sigma}\right) dt = a^* = \int_{-x}^{x} f(t) dt$ (D.1 step 2), then $x_\sigma = \sigma x$.

According to the Substitution theorem of Riemann Integral, for any $g: [a,b] \to \mathbf{R}$ continuously differentiable on (a, b), If $f: I \to \mathbf{R}$ is continuous on an interval $I \supset g([a,b])$, then

$$\int_a^b f(g(x)) g'(x) dx = \int_{g(a)}^{g(b)} f(t) dt$$

For our case, if we take $g(x) = \frac{x}{\sigma}$, then we have

$$a^* = \frac{1}{\sigma} \int_{-x_\sigma}^{x_\sigma} f\left(\frac{t}{\sigma}\right) dt = \int_{-x_\sigma}^{x_\sigma} f\left(\frac{t}{\sigma}\right) \frac{1}{\sigma} dt = \int_{-\frac{x_\sigma}{\sigma}}^{\frac{x_\sigma}{\sigma}} f(y) dy$$

Moreover, $a^* = \int_{-x}^{x} f(t) dt$. Comparing it with the above equation we obtain $\frac{x_\sigma}{\sigma} = x$ or equivalently $x_\sigma = \sigma x$. Now let $n$ be found from Step 3 of Algorithm D.1. Therefore, n is such that

$$\int_{-(2n+1)x}^{(2n+1)x} f(x) dx \geq 1 - \tau^*, \quad \text{but}$$

$$\int_{-(2n-1)x}^{(2n-1)x} f(t) dt < 1 - \tau^*$$

$$\int_{-(2n+1)x_\sigma}^{(2n+1)x_\sigma} \frac{1}{\sigma} f\left(\frac{t}{\sigma}\right) dt = \int_{-(2n+1)\sigma x}^{(2n+1)\sigma x} \frac{1}{\sigma} f\left(\frac{t}{\sigma}\right) dt$$
$$= \int_{-(2n+1)x}^{(2n+1)x} f(t) dt \geq 1 - \tau^*$$

Similarly we can show that $\int_{-(2n-1)x_\sigma}^{(2n-1)x_\sigma} \frac{1}{\sigma} f\left(\frac{t}{\sigma}\right) dt < 1 - \tau^*$. Therefore n is the smallest natural number which satisfies $\int_{-(2n+1)x}^{(2n+1)x} \frac{1}{\sigma} f\left(\frac{t}{\sigma}\right) dt \geq 1 - \tau$ and hence $n_\sigma = n$.

For the 2nd part of the proposition, we notice that $a_j(\sigma) = \int_{(2j-1)x_\sigma}^{(2j+1)x_\sigma} \frac{1}{\sigma} f\left(\frac{t}{\sigma}\right) dt = \int_{(2j-1)\sigma x}^{(2j+1)\sigma x} \frac{1}{\sigma} f\left(\frac{t}{\sigma}\right) dt = \int_{(2j-1)x}^{(2j+1)x} f(t) dt = a_j$ for $l \in \{-n, \ldots, n\} \setminus \{0\}$

From initial condition we have $a^* = a_l = a_l(\sigma)$, and hence $a_l = a_l(\sigma) \, \forall \, l \in \{-n, \ldots, n\}$, that is the same weight vector in either case. ∎

## III. APPLICATION TO NEURAL SIGNALS

### A. Data

ECoG data of 21 epileptic patients containing 87 focal onset seizures have been obtained from the Freiburg Seizure Prediction Project (https://epilepsy.uni-freiburg.de/2008). One hour recording containing preictal, ictal and postictal ECoG of 1 h duration in each of the 87 cases is available. The ECoG data were acquired using Neurofile NT digital video EEG system (It-med, Usingen, Germany) with 128 channels, 256 Hz sampling rate, and a 16 bit analog to digital converter. In all cases the ECoG from only six sites have been analyzed, because only six channel data were made available through the above website. However this is a publicly available data set and therefore good for benchmarking novel algorithms. Three of the six channel data are from the focal areas and the other three from outside the focal areas. For each patient there are 2–5 h of ictal data (actually preictal + ictal + postictal) recordings. Each hour's recording contains only one seizure of few tens of seconds to a couple of minute duration.

### B. Results

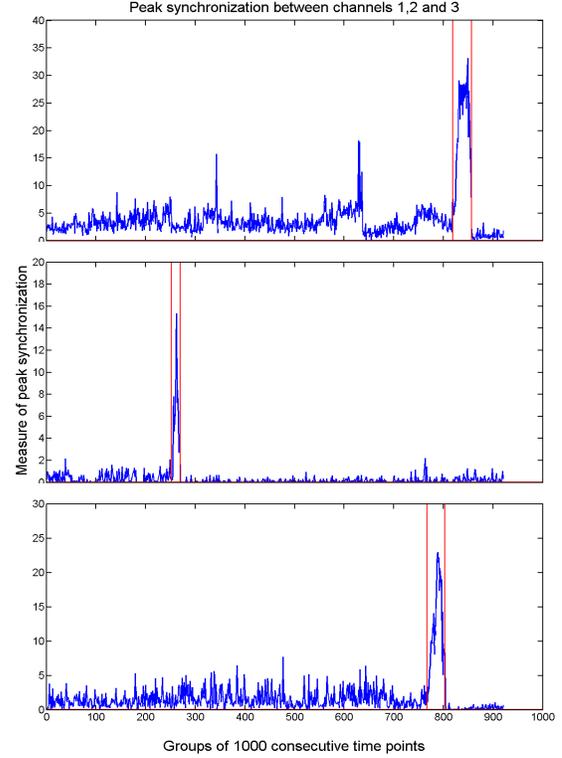

Fig 4. Each of the above subplots depicts the measure of peak synchronization of neural signals from the 3 focal channels of patient 2. The 3 subplots correspond to all the three different hours of recording from top to bottom respectively. The vertical red lines mark the seizure onset and offset. The horizontal red line marks the line of statistical significance. Seizure part is clearly distinguishable by abnormally high synchronization values.

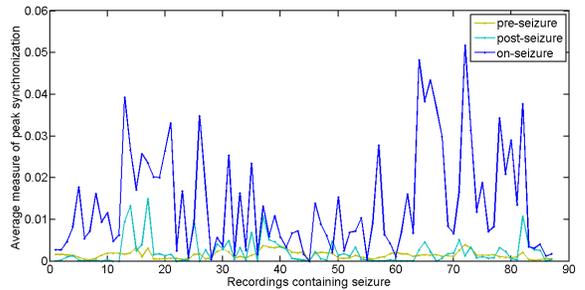

Fig 5. The above figure depicts 3 plots, points in which indicate the average measure of peak synchronization of neural signals from the 3 focal channels for pre (red), post (green) and during (blue) seizure for the 87 recordings (arranged along abscissa). It clearly shows the trend of excessive peak-synchronization during seizures, compared to before or after for most of the seizures considered.

Table 1. Summary of results – Peak synchronization
A total of 87 seizures recorded from the focal ECoG of 21 focal epileptic patients have been analyzed.

| | | |
|---|---|---|
| Total number of seizures out of 87 seizures for which peak synchronization during seizure is very high.<br><br>High value is desirable[2]. | 78 | 89.66% |
| Number of seizures for which peak synchronization pre-seizure is more than during seizure.<br><br>Low value is desirable. | 6 | 6.90% |
| Number of seizures for which peak synchronization post-seizure is more than during seizure.<br><br>Low value is desirable. | 3 | 3.45% |

For peak detection, a standard amplitude thresholding (median + 2 times standard deviation [42]) technique was applied to the band-pass filtered and notched filtered signals within 25-100 Hz, with 49-51 Hz notching frequency. We have also checked by trial and error that this threshold worked well for epileptic spike identification.

The peak synchronization measure was found using the Gaussian function with mean 0 and scale parameter 1 as the weight determination function, $a_0 = 0.5$ and threshold, $\tau = 0.001$. The proposed measure indicates seizure, by showing very high value during seizure (See Fig 4 and Fig 5), which complies with the definition of epileptic seizure proposed by the International League Against Epilepsy (ILAE) and the International Bureau for Epilepsy (IBE), "An epileptic seizure is a transient occurrence of signs and/or symptoms due to abnormal excessive or synchronous neuronal activity of the brain'' [43]. Immediately after the seizure-offset, the measure is greatly diminished. We also recorded the average measures across all three focal channel pairs (= peak synchronization across those three channels) for 21 patients (a total of 87 seizures), during seizure and considering 25000 time points ($\approx$ 98s) pre and post seizure (See Table 1). 78 out of 87 seizures showed very high measure of peak synchronization during seizure, compared to before and after it. That is, there are only 9 exceptions out of 87 cases or 10.34%.

We have also tested the statistical significance of the outcome of peak synchronization measure. Since here we are concerned about synchronization among three channels, we have taken 100 triplets of shifted surrogate signals generated by randomly shuffling the real signal amplitude across the time (Clearly, generated signals are of same length and amplitude as the real signals). Then we have determined speak synchronization measure for each of the generated triplets. The value of statistical significance has been chosen to be the value that is above 95% of the peak synchronization measure values of the 100 shifted surrogate triplets (Horizontal red line in Fig 4). Peak synchronization value above this value (above the red line in Fig 4) signifies that the synchronization is not due to mere chance.

IV. COMPARISON WITH AMPLITUDE CORRELATION

Synchronization across channels before, during and after an epileptic seizure has been analyzed in many different ways. Enhanced phase synchronization across focal areas during seizure has been observed in [4]. Multi-channel amplitude correlation among focal channels during seizure has been studied in [44], [45] in order to understand the seizure dynamics. Since the applied correlation measure in [44] and [45] is multi-channel, it would be more appropriate to compare the peak synchronization measure with this multi-channel correlation measure.

If there are r number of channels, then a r × r cross-correlation matrix has to be formed. The matrix is calculated for cross-correlation over a window with m time points. Then r eigen values of the matrix are calculated and sorted in descending order. Then the window is slided (usually continuously) and the process is repeated. The temporal plot of the highest eigen value is generated by the highest eigen values at all time points. If the highest eigen value plot is increasing with respect to time, it is said that the overall amplitude correlation is growing up. If it is decreasing, the overall correlation is also decreasing (Fig. 6). For more detail see [44] and [45].

Table 2: Summary of results – Amplitude Correlation

| | | |
|---|---|---|
| Total number of seizures out of 87 seizures for which amplitude correlation during seizure is higher than pre- or post-seizure.<br><br>High value is desirable. | 24 | 27.59% |
| Number of seizures for which amplitude correlation before seizure is more than during seizure.<br><br>Low value is desirable. | 36 | 41.38% |
| Number of seizures for which amplitude correlation is more after seizure than during seizure.<br><br>Low value is desirable. | 63 | 72.41% |

In Fig. 7, we have plotted the average amplitude correlation among three focal channels before (25000 time points as in case of peak synchronization), during and after (25000 time points as in case of peak synchronization) seizure. In Fig. 7 different graphs signifying correlation measure before (red), during (blue) and after (green) are lot more overlapping than the graphs in Fig. 5. Thus a clear trend of high synchronization during an epileptic seizure is not that evident in Fig. 7 as it is in Fig. 5. The same becomes evident comparing between Table 1 and Table 2. Thus we can conclude that the notion of hyper-synchronous neuronal activity associated with epileptic seizures is modeled better by peak synchronization than amplitude correlation.

---
[2] See the definition of epileptic seizure by ILAE and IBE (ref. [43]), mentioned earlier.

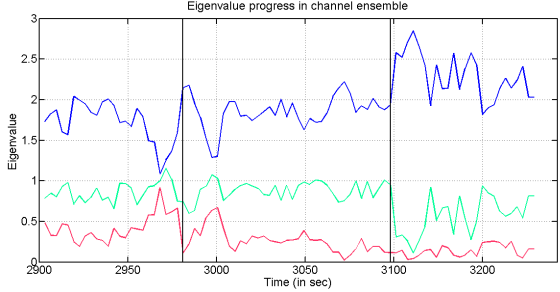

Fig. 6. Three eigenvalues of the amplitude correlation matrix during the 21st hour of recording of patient 2. Vertical lines indicate seizure onset and offset points.

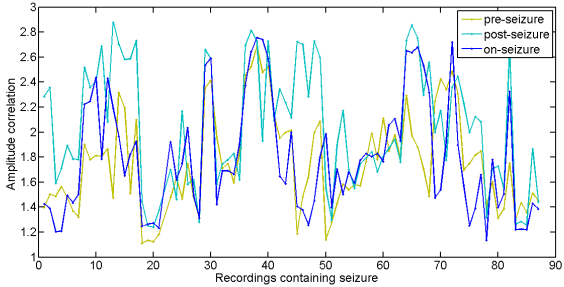

Fig. 7. The above figure depicts 3 lines, points in which indicate the amplitude correlation of neural signals from the 3 focal channels pre, post and during seizure for the 87 recordings containing seizure.

In fact peak synchronization is a more appropriate measure of synchronization among ECoG signals during focal epilepsy than amplitude correlation or phase synchronization, because peaks (spikes) are distinct features of ECoG signals during an epileptic seizure. Sporadic sharp spikes occur in ECoG of epileptic patients even when there is no seizure (for example, in between two successive seizures, which is called interictal period). ECoG spikes are created due to simultaneous (synchronous) firing of a large number of neurons in a small neighborhood of the channel. Focal channels are all in seizure onset zone and they have spikes in temporal proximity of one another. This happens due to heterogeneous spread of seizures from the focal points [46]. This is an ideal situation where the peak synchronization measure is able to capture the underlying synchronization more fully than amplitude correlation.

## V. CONCLUSION

In many applications a higher-than-background amplitude peak in a time domain pure signal indicates an important event. If two pure signals are mutually dependent, then occurrence of a peak in one should be dependent on occurrence of peaks in the other. This dependence of peaks of one signal on the peaks of the other has been modeled for the first time in this paper, which gives an event-based mutual dependence measure between the two signals. We have shown that the proposed peak synchronization measure is largely independent of the different choices of values for the model parameters. We subsequently generalized this measure for more than two signals. To the best of our knowledge the two important concepts of signals namely, peak and synchronization, have been put together for the first time in this work to define a peak synchronization measure among multiple signals. One important advantage of the algorithm, that we proposed in this work to measure the synchronization, is its ability to measure synchronization even when peaks are occurring with some time lag across different channels. Most other measures of synchronization (such as Hilbert [8] or wavelet [47] transformation based phase synchronization) lack the ability to take care of the time lag in occurrence of near simultaneous events across different channels.

Next, we showed the effectiveness of the proposed measure on a large biomedical signal database. This database was available publicly from https://epilepsy.uni-freiburg.de/2008, where we downloaded it from in 2009, free of charges and it was specifically meant for benchmarking new algorithms. Now this database has been merged with a much bigger dataset available for a price. Please visit http://epilepsy-database.eu/ and see ref. [48] for more detail. Our assertion was the newly developed peak synchronization measure is modeling the synchronization more comprehensively than classical amplitude correlation measure. The amplitude correlation measure was applied earlier to study same type of synchronization in similar signals [44], [45], but with suboptimal outcome. In fact evolution of synchronization across different channels during an epileptic seizure is a complex process. Synchronization is less at the onset of seizure, but more at the offset and even beyond as measured by phase synchronization and amplitude correlation [49], [50]. Quantitative outcomes of the two methods differ significantly and even some time contradicted each other (see ref. [49], [50] for detail). In this scenario peak synchronization measure has been able to establish higher synchronization during seizure than before or after it more decisively on our dataset with only about 10% exception. In future we have plans to apply this method to study interictal discharges in scalp EEG of epileptic patients during sleep. During sleep, scalp EEG usually contains lesser artifacts than wake state EEG, which will make detection of peaks of neural origin more accurate.

One distinct advantage of the peak synchronization measure algorithm is its linear time complexity (see Appendix A). For $r$ many signals, each $N$ time point long, the algorithm takes $O(r^2 N)$ time to execute. This makes the algorithm suitable for online implementation.

However peaks, both upright and inverted taken together, do not in general constitute the information content of a signal. In this respect Fourier coefficients are more reliable components of a signal. In future development of the measure peaks may be replaced by Fourier coefficients and the measure may be calculated in the frequency domain rather than in the time domain. This will make the peak synchronization measure much closer to mutual information measure. Even then the present form of the peak synchronization will remain useful for signals, in which time domain peaks contain important information (such as biomedical signals).

The proposed peak synchronization measure may find applications in a number of areas. For example, Fig 4 and Fig 5 indicate the potential of the measure for automatic seizure detection. Also hemodynamic correlate of interictal spikes may be better related to spikes occurring near simultaneously across several channels of either scalp or depth EEG rather than spikes occurring in a single channel. A study on how the hemodynamic response function varies with the degree of synchronization across the event over a period of time can give us a new insight into the seizure dynamics. In cognitive science research a study on how event related potential (ERP) peaks across different channels synchronize with respect to the stimulus presentation may supplement our knowledge of neuronal synchronization leading to ERP generation. This may be useful in brain computer interface.

ACKNOWLEDGMENT

RB and KK want to thank the Dean of Studies, Indian Statistical Institute and the institute's Bangalore Center for generous hospitality and support during the work. RB is recipient of an Innovation in Science Pursuit for Inspired Research (INSPIRE) fellowship from the Department of Science and Technology (DST), Government of India. KK is recipient of Kishore Vaigyanik Protsahan Yojana (KVPY) scholarship from the DST. We also thank the three anonymous reviewers for helpful comments leading to improvement of the manuscript.

APPENDIX A

I. PSEUDOCODE FOR MEASURE CALCULATION AND COMPLEXITY ANALYSIS:

*A. Finding the Coefficient Vector*

**Input:** *th*, denoting the threshold after which tail areas (described previously) are considered to be insignificant.
$ca$, the central coefficient.
**Output:** CV, the coefficient vector.

- $normcdf(x, \mu, \sigma)$ -> computes univariate normal distribution function at x, corresponding to mean $\mu$ and variance $\sigma^2$.
- $norminv(P, \mu, \sigma)$ -> computes inverse of univariate normal distribution function with mean $\mu$ and variance $\sigma^2$.
- horzcat (A, B) -> horizontally concatenates two matrices A and B to form a single matrix.
- taking $\sigma = 1$.

**begin**
(1)   $gap := 2 * norminv((ca + 1)/2, 0, \sigma)$
(1)   $start := gap/2$
(1)   $a := (ca + 1)/2$
(1)   $b := ca$

(2)   while $(1 - a) > th$
(2)   $c := normcdf(start + gap, 0, \sigma) - a$
(2)   $a := normcdf(start + gap, 0, \sigma)$
(2)   $b := horzcat(b, c)$
(2)   $start := start + gap$
(2)   end

(3)   $CV := horzcat(b(length(b): -1: 2), b)$

**end**

*B. Calculating the measure of peak synchronization using the above found coefficient vector*

**Input:** A, an $r \times N$ matrix, consisting of r many peak detected signals, with N time points each.

CV, the coefficient vector

**Output:** mc, a $1 \times N$ vector, with each index carrying the measure of peak synchronization for the r signals at that time point.

**begin**
(1)   F: = $r \times N$ matrix of zeroes
(1)   n: = floor($length(CV)/2$)
(1)   S: = $\binom{r}{2} \times N$ matrix of zeroes

(2)   for $i := 1 : r$
(2)   for $t := n + 1 : N - n$
(2)   $F(i, t) := A(i, (t - n: t + n)) \times CV^T$
(2)   end
(2)   end

(3)   $z := 1$
(3)   for $i := 1 : r$
(3)     for $j := i + 1 : r$
(3)       for $t := n + 1: N - n$
(3)         if $(A(i, t) \times A(j, t) = 0)$
(3)           $S(z, t) := max[ (F(i, t) \times A(j, t)), (F(j, t) \times A(i, t)) ]$
          else
(3)           $S(z, t) := (F(i, t) \times A(j, t) + F(j, t) \times A(i, t))/2;$
          end
        end
(3)   z=z+1;
end
end

(4)    mc=mean(rows of S);

**end**

*C. Complexity of the above pseudocode:*

For a fixed $a_0$, part A takes constant time, being free of r or N and hence a constant function of them.

In part B, F consists of $(f_1, \ldots, f_r)$ as its rows, with each $f_i$, taking values over N time points.
S contains $\phi_{i,j}$s as its rows for $i \neq j$, with $\phi_{i,j}$ representing the measure of peak synchronization between the $i^{th}$ and $j^{th}$ signals over N time points.
CV is the coefficient vector of length 2n+1.

Step (1) deals with pre-allocation of F and S, and assignment of n. In step (2) the matrix F, or equivalently the $f_i$s, is calculated. This takes time proportional to $r(N - 2n)$. In step (3), the matrix S, or equivalently the pairwise measures of peak synchronization, as discussed previously, is calculated. Time taken by this step is at most $cr(r - 1)(N - 2n)$, for some constant $c$. Step (4) finally calculates the measure for r many signals, taking time proportional to $rN$.

The most expensive step in the above algorithm is calculating the S matrix. So the run time of Part B of the algorithm is $O(r(r - 1)(N - 2n))$, and hence $O(r^2 N)$. In most practical cases, r even has a finite upper bound which does not depend on $N$, in which case, the time complexity is $O(N)$.

**Rahul Biswas** was born in December 1992. He has completed his Bachelor of Statistics, 3rd year from the Indian Statistical Institute (ISI), Kolkata in May 2014. He has worked with different centers of ISI, with the Tata Institute of Fundamental Research (TIFR), India, and Johns Hopkins Bloomberg School of Public Health, USA. His research interests include fundamental statistical inference, non-parametric statistics, signal processing, probability theory, high dimensional statistics and biostatistics.

Mr. Biswas is recipient of the INSPIRE scholarship from the Department of Science and Technology (DST), Government of India, and earlier received the National Initiative for Undergraduate Science (NIUS) fellowship from the Tata Institute of Fundamental Research (TIFR), India, among several other awards and honors. He has secured top ranks in many nationwide entrance exams and was an awardee of the National Standard Examination in Physics, known as the Regional Physics Olympiad, and also of the Regional Mathematics Olympiad.

**Koulik Khamaru** was born in March 1994 and has completed his Bachelor of Statistics, 3rd year from the Indian Statistical Institute (ISI), Kolkata in May 2014. He has worked in different institutes including ISI Bangalore Center and Columbia University. He was selected for numerous national awards and fellowships including INSPIRE and KVPY fellowships of the DST, Government of India, Jagadis Bose National Science Talent Search scholarship and Certificate of Merit in Indian National Mathematics Olympiad. Apart from ISI he secured admission in quite a few top Indian institutes for higher studies including Chennai Mathematical Institute and Indian Institute of Science. His interests are in statistical modeling and inference, signal processing, probability theory, bio statistics, multivariate analysis, linear algebra and analytical properties of linear spaces.

Mr. Khamaru is recipient of KVPY fellowship from the DST.

**Kaushik K. Majumdar** (M'06–SM'10) was born in January 1967 and received the B.Sc. (Pass) and Mathematics Honours (bridge course) degrees from the University of Calcutta, Calcutta, India, in 1987 and 1990 respectively. He received M.Sc. in Mathematics from the Annamalai University, Annamalai Nagar, India, 1996, and the M.Tech. and Ph.D.degrees in Computer Science from the Indian Statistical Institute, Calcutta, India, in 1999 and 2003, respectively.

He worked in University of Memphis, University of Oregon and Florida Atlantic University, all in USA. He also worked in INRIA Sophia Antipolis in France. He was an R & D Scientist in Electrical Geodesics Inc. (EGI) Eugene, Oregon, USA. Currently he is an Associate Professor in the Indian Statistical Institute, Bangalore Center, where he heads the Computational Neuroscience Group. His interests are in neural signal and information processing (EEG, ECoG, LFP, fMRI), neural information coding and mathematical and computational modeling of the brain functions.

Dr. Majumdar is a member of the American Mathematical Society and the Society for Neuroscience.